\documentclass[10pt, conference, compsocconf]{IEEEtran}
\usepackage[T1]{fontenc}
\usepackage[latin9]{inputenc}
\usepackage{array}
\usepackage{multirow}
\usepackage{cite}
\usepackage{amsmath,amssymb,amsfonts}
\usepackage{algorithmic}
\usepackage{graphicx}
\usepackage{textcomp}
\usepackage{xcolor}
\begin{document}

\title{A Fuzzy Post-project Evaluation Approach for Security Video Surveillance System}

\author{\IEEEauthorblockN{1\textsuperscript{st} Ming Liu}
\IEEEauthorblockA{\textit{School of Electronics, Information} \\
\textit{and Electrical Engineering}\\
\textit{Shanghai Jiao Tong University}\\
Shanghai, China \\
liuming198904@sjtu.edu.cn}
\and
\IEEEauthorblockN{2\textsuperscript{nd} Zhi Xue}
\IEEEauthorblockA{\textit{School of Electronics, Information} \\
	\textit{and Electrical Engineering}\\
	\textit{Shanghai Jiao Tong University}\\
Shanghai, China \\
zxue@sjtu.edu.cn}

}

\maketitle

\begin{abstract}
Video surveillance is an essential component of the public security system. The security video surveillance system is a powerful means to prevent violence and crimes, and it is closely coupled with the construction of smart cities. A post-project evaluation is an evaluation of a project's actions and outcomes after its completion. Post-project evaluation can scientifically and objectively evaluate the construction effectiveness of video surveillance system at a certain stage. Utilizing post-project evaluation can find out the causes of success or failure to make recommendations for the construction of a security video surveillance system in the next stage. Therefore, we propose a fuzzy post-project evaluation approach for the security video surveillance system in a real-world community. The fuzzy theory and fuzzy multi-level evaluation method are applied. The evaluation result demonstrates that the proposed approach is practically applicable to real-world security video surveillance systems.
\end{abstract}


\section{Introduction}
Video surveillance is an essential component of the public security system. The traditional video surveillance system includes front-end cameras, transmission cables, and video surveillance platforms\cite{Collins2000}. In recent years, with the rapid development of hardware and artificial intelligence technologies, video surveillance technologies have also reached another level\cite{Vishwakarma2013}\cite{Ojha2015}. The construction and development of security video surveillance systems contribute to various fields such as crime detection, urban management, environment protection, risk monitoring, emergency response, and other security issues. The security video surveillance system plays a vital role in the aspect of investigation and case solving, and it is a powerful means to prevent violence and crimes\cite{TDraty2010}. Video surveillance is tightly coupled with the construction of smart cities. 

A post-project evaluation (or post-project review)\cite{Anbari2008Post} is an evaluation of a project's actions and outcomes after its completion. A post-project evaluation should be performed within one year after completion of the project. A post-project evaluation acknowledges a project's accomplishments and claims people's work. It recognizes effective methods and strategies and ensures they can be utilized in the future. It also recognizes ineffective practices, and ensures they will not be applied in the future. Post-project evaluation can scientifically and objectively evaluate the construction effectiveness of video surveillance system at a particular stage. The general objective of the post-project evaluation is to systematically analyze the purpose, the implementation process, and the impact of the completed project. The post-project evaluation determines whether the project's expected goal has been achieved and whether the planning is reasonable. The post-project evaluation also finds out the causes of success or failure to make recommendations for the future decision of new projects. Therefore, utilizing the post-project evaluation approach helps the construction of the security video surveillance system in the next stage.

To the best of our knowledge, when it comes to the evaluation of a security video surveillance system, currently, there is no paper using an effective index system specially designed for it. Motivated by this issue, we contribute a fuzzy post-project evaluation approach for the security video surveillance system in a real-world community. Considering that many evaluation indexes have certain fuzziness, the fuzzy theory and fuzzy multi-level evaluation method are used. Meanwhile, the evaluation methods include seeking advice from consultants and distributing questionnaires.

The rest of this paper is organized as follows. Section \ref{index system} introduces the establishment of the index system and comment set. Section \ref{evaluation process} provides the evaluation process. Section \ref{Conclusion} provides the conclusion and future work.

\section{Establishment of the index system and comment set}
\label{index system}
\subsection{Symbols and descriptions}
Symbols used in the proposed approach are listed and described in table \ref{Symbols}.
\begin{table}[htbp]
	\centering
	\caption{Symbols and descriptions}
	\begin{tabular}{|>{\raggedright}p{0.9cm}|>{\raggedright}p{7cm}|}
		\hline 
		Symbols & Descriptions\tabularnewline
		\hline 
		\hline 
		$A$ & The weight vector of $U$\tabularnewline
		\hline 
		$B$ & The evaluation result vector of $U$\tabularnewline
		\hline 
						$M$ & The membership of a secondary index\tabularnewline
		\hline 
				$U$ & The index set\tabularnewline
		\hline 
		$U'$ & The secondary layer of the index set\tabularnewline
		\hline 
		$V$ & The comment set\tabularnewline
		\hline 
						$a$ & The weight vector of $u$\tabularnewline
		\hline 
		$b$ & The evaluation result vector of $u$\tabularnewline
		\hline 
		$m$ & The number of valid feedback for a secondary index that commented as $v$ \tabularnewline
		\hline 
		$n$ & The number of valid feedback\tabularnewline
		\hline 
		$r$ & The evaluation matrix of a primary index\tabularnewline
		\hline 
		$u$ & A single index\tabularnewline
		\hline 
		$v$ & A single comment\tabularnewline
		\hline 
	\end{tabular}
	\label{Symbols}
\end{table}

\subsection{Establishment of the index system}
The establishment of an index system is the foundation of post-project evaluation. The index system should be composed of several independent indexes. Theoretical analysis and expert consultation are two commonly used methods to select and adjust indexes. The following principles need to be considered when constructing the index system:
\begin{enumerate}
\item Integrity. The index system should be detailed and appropriate to describe the whole picture of the project.
\item Operability. Each index should accurately establish specific content and meaning.
\item Systematicness. The index system should be hierarchical; each evaluation index should be independent yet related to each other. The index system should describe the features and states of the video surveillance system from multiple views.
\item Cost-effectiveness. The index system should evaluate the whole process of the project with relatively less time and labor costs.
\item Guidance. The purpose of the post-project evaluation is to evaluate the advantages and disadvantages and encourage the project to develop in the correct direction. This principle needs to be considered when constructing the index system.
\end{enumerate}

\begin{table}[tbp]
	\centering
	\caption{The index system\cite{wangyongsheng}}
	\begin{tabular}{|>{\raggedright}m{2.5cm}|>{\raggedright}m{5.1cm}|}
		\hline 
		Primary Indexes & Secondary Indexes\tabularnewline
		\hline 
		\hline 
		\multirow{2}{3cm}{Information Resources\\($u_1$)} & Resources Opening ($u_{11}$)\tabularnewline
		\cline{2-2} 
		& Resources Sharing ($u_{12}$)\tabularnewline
		\hline 
		\multirow{5}{3cm}{Information Security\\($u_{2}$)} & Physical Security ($u_{21}$)\tabularnewline
		\cline{2-2} 
		& Network Security ($u_{22}$)\tabularnewline
		\cline{2-2} 
		& System Security ($u_{23}$)\tabularnewline
		\cline{2-2} 
		& Operation Security ($u_{24}$)\tabularnewline
		\cline{2-2} 
		& Data Security ($u_{25}$)\tabularnewline
		\hline 
		\multirow{3}{3cm}{Mechanism Guarantee\\($u_3$)} & Planning And Construction Schemes ($u_{31}$)\tabularnewline
		\cline{2-2} 
		& Organizational Management Mechanism ($u_{32}$)\tabularnewline
		\cline{2-2} 
		& Technical Support Capability ($u_{33}$)\tabularnewline
		\hline 
		\multirow{4}{3cm}{Operation and Maintenance\\($u_4$)} & Operation and Maintenance Mechanism ($u_{41}$)\tabularnewline
		\cline{2-2} 
		& Operation and Maintenance Team ($u_{42}$)\tabularnewline
		\cline{2-2} 
		& Operation and Maintenance Funds ($u_{43}$)\tabularnewline
		\cline{2-2} 
		& Operation and Maintenance Quality ($u_{44}$)\tabularnewline
		\hline 
		\multirow{5}{3cm}{Infrastructure\\($u_5$)} & Video Monitoring Coverage($u_{51}$)\tabularnewline
		\cline{2-2} 
		& Video Monitoring Networking Rate ($u_{52}$)\tabularnewline
		\cline{2-2} 
		& HD Video Construction Rate ($u_{53}$)\tabularnewline
		\cline{2-2} 
		& Video Equipment Integrity ($u_{54}$)\tabularnewline
		\cline{2-2} 
		& Video Equipment Archive Rate ($u_{55}$)\tabularnewline
		\hline 
		\multirow{4}{3cm}{Public Security Application\\($u_6$)} & Construction of Investigation Team ($u_{61}$)\tabularnewline
		\cline{2-2} 
		& Construction of Video Analysis Platform ($u_{62}$)\tabularnewline
		\cline{2-2} 
		& Case Investigation Assistance Rate ($u_{63}$)\tabularnewline
		\cline{2-2} 
		& Suspects Arresting Assistance Rate ($u_{64}$)\tabularnewline
		\hline 
		\multirow{4}{3cm}{Comprehensive Administration Application\\($u_7$)} & Construction of Comprehensive Administration Center ($u_{71}$)\tabularnewline
		\cline{2-2} 
		& Construction of Video Sharing Platform ($u_{72}$)\tabularnewline
		\cline{2-2} 
		& Urban Management Application  ($u_{73}$)\tabularnewline
		\cline{2-2} 
		& Comprehensive Governance Application  ($u_{74}$)\tabularnewline
		\hline 
		\multirow{2}{3cm}{Information Services\\($u_8$)} & Construction of Shared Service Platform ($u_{81}$)\tabularnewline
		\cline{2-2} 
		& Video Monitoring Resource Sharing ($u_{82}$)\tabularnewline
		\hline 
		\multirow{2}{3cm}{Public Experience\\($u_9$)}& User Experience Survey ($u_{91}$)\tabularnewline
		\cline{2-2} 
		& Public Experience Survey ($u_{92}$)\tabularnewline
		\hline
	\end{tabular}
	\label{The index system}
\end{table}
We refer to the index system developed by Wang et al. \cite{wangyongsheng} in this paper. This index system is divided into two levels (primary indexes and secondary indexes). The index system is shown in table \ref{The index system}. Table \ref{The explanation of index system} explains the indexes. For index $u_{91}$ and $u_{92}$, questionnaires are assigned to the system users and the public. The consultants will provide comments based on the survey results.

\begin{table}[t]
	\centering
	\caption{The explanation of indexes\cite{wangyongsheng}}
	\begin{tabular}{|>{\raggedright}p{0.8cm}|>{\raggedright}p{6.9cm}|}
		\hline 
		Indexes & Explanation\tabularnewline
		\hline 
		\hline 
		$u_{11}$ & The scope of video surveillance information that open to the community\tabularnewline
		\hline 
		$u_{12}$ & Application effect of video information-sharing mechanism\tabularnewline
		\hline 
		$u_{21}$ & Operation reliability of system hardware equipment\tabularnewline
		\hline 
		$u_{22}$ & Soundness of network security management mechanism\tabularnewline
		\hline 
		$u_{23}$ & Security capability of the information system related to video surveillance\tabularnewline
		\hline 
		$u_{24}$ & Capability to ensure the stability and security of information system
		during operation\tabularnewline
		\hline 
		$u_{25}$ & Capability to ensure the security of video data during collection,
		storage, transmission, etc.\tabularnewline
		\hline 
		$u_{31}$ & The quality of planning and construction scheme\tabularnewline
		\hline 
		$u_{32}$ & Soundness of organizational management mechanism\tabularnewline
		\hline 
		$u_{33}$ & Technical support capability of system construction, operation, management,
		etc.\tabularnewline
		\hline 
		$u_{41}$ & Soundness of system operation and maintenance mechanism\tabularnewline
		\hline 
		$u_{42}$ & Professionalization of system operation and maintenance team\tabularnewline
		\hline 
		$u_{43}$ & Guarantee of system operation and maintenance funds\tabularnewline
		\hline 
		$u_{44}$ & The effectiveness of quality supervision mechanism for system operation
		and maintenance\tabularnewline
		\hline 
		$u_{51}$ & Coverage of video surveillance in key public areas\tabularnewline
		\hline 
		$u_{52}$ & Networking rate of video surveillance in key public areas\tabularnewline
		\hline 
		$u_{53}$ & Construction of HD video surveillance in key public areas\tabularnewline
		\hline 
		$u_{54}$ & Integrity of video surveillance in key public areas\tabularnewline
		\hline 
		$u_{55}$ & Soundness of video surveillance equipment management system\tabularnewline
		\hline 
		$u_{61}$ & The effectiveness of the construction of video surveillance application
		team\tabularnewline
		\hline 
		$u_{62}$ & Application of video information based on new technology\tabularnewline
		\hline 
		$u_{63}$ & The contribution of video surveillance to the detection of criminal
		cases\tabularnewline
		\hline 
		$u_{64}$ & The contribution of video surveillance to public security operations\tabularnewline
		\hline 
		$u_{71}$ & Ability to provide quality public management services\tabularnewline
		\hline 
		$u_{72}$ & The application of video information in urban and rural governance\tabularnewline
		\hline 
		$u_{73}$ & Application effect of video surveillance in public order maintenance\tabularnewline
		\hline 
		$u_{74}$ & Application effect of video surveillance in crowded places\tabularnewline
		\hline 
		$u_{81}$ & The ability of aggregation, sharing, and access control for video
		information\tabularnewline
		\hline 
		$u_{82}$ & The ability to provide video information to users through
		public video sharing platform\tabularnewline
		\hline 
		$u_{91}$ & The satisfaction of system users for the construction of video surveillance system; questionnaires are assigned to the system users\tabularnewline
		\hline 
		$u_{92}$ & The satisfaction of the public for the construction of video surveillance system; questionnaires are assigned to the public
		\tabularnewline
		\hline 
	\end{tabular}
	\label{The explanation of index system}
\end{table}

\subsection{The definition of comment set and index set}
In this paper, five class fuzzy language is used to define the comment set. The five classes are "Excellent (Level 5), Good (Level 4), Average (Level 3), Fair (Level 2), and Poor (Level 1)". Therefore, the comment set is:
\begin{flalign}\nonumber
&V=\{v_1, v_2, v_3, v_4, v_5\} = \{5, 4, 3, 2, 1\}&
\end{flalign}
\indent As two levels are used for evaluation, the main evaluation system is divided into 9 subsystems, and 31 single indexes related to the subsystems are selected. Therefore, the index set is divided into two layers, the primary layer $U$, and the secondary layer $U'$. The primary layer $U$ is:
\begin{flalign}\nonumber
&U=\{u_1, u_2, u_3, u_4, u_5, u_6, u_7, u_8, u_9\}&
\end{flalign}
The secondary layer $U'$ is:
\begin{flalign}\nonumber
&U'=\left\{
\begin{aligned}
&u_1=\{u_{11},u_{12}\}\\
&u_2=\{u_{21},u_{22},u_{23},u_{24},u_{25}\}\\
&u_3=\{u_{31}, u_{32}, u_{33}\}\\
&u_4=\{u_{41},u_{42},u_{43},u_{44}\}\\
&u_5=\{u_{51},u_{52},u_{53},u_{54},u_{55}\}\\
&u_6=\{u_{61},u_{62},u_{63},u_{64}\}\\
&u_7=\{u_{71},u_{72},u_{73},u_{74}\}\\
&u_8=\{u_{81},u_{82}\}\\
&u_9=\{u_{91},u_{92}\}
\end{aligned}
\right.&
\end{flalign}
\begin{table}[t]
	\caption{The weight table}
	\label{Weight table}
	\centering
	\begin{tabular}{|l|l|l|l|l|l|l|l|l|l|}
		\hline
		\multicolumn{3}{|l|}{\multirow{2}{*}{Hierarchy}}                & \multicolumn{5}{l|}{Consultant Weights} & \multirow{2}{*}{Avg.} & \multirow{2}{*}{Norm.} \\ \cline{4-8}
		\multicolumn{3}{|l|}{}                                          & (1)    & (2)    & (3)   & (4)   & (5)   &                       &                        \\ \hline
		\multicolumn{2}{|l|}{\multirow{9}{*}{\rotatebox{90}{Primary Indexes}}}    & $u_1$  & 1      & 1      & 1     & 1     & 1     & 1                     & 0.1                    \\ \cline{3-10} 
		\multicolumn{2}{|l|}{}                                    & $u_2$  & 2      & 1      & 1     & 1     & 1     & 1.2                   & 0.12                   \\ \cline{3-10} 
		\multicolumn{2}{|l|}{}                                    & $u_3$  & 1      & 1      & 1     & 1     & 1     & 1                     & 0.1                    \\ \cline{3-10} 
		\multicolumn{2}{|l|}{}                                    & $u_4$  & 1      & 2      & 1     & 1     & 1     & 1.2                   & 0.12                   \\ \cline{3-10} 
		\multicolumn{2}{|l|}{}                                    & $u_5$  & 1      & 1      & 2     & 1     & 1     & 1.2                   & 0.12                   \\ \cline{3-10} 
		\multicolumn{2}{|l|}{}                                    & $u_6$  & 1      & 1      & 1     & 2     & 1     & 1.2                   & 0.12                   \\ \cline{3-10} 
		\multicolumn{2}{|l|}{}                                    & $u_7$  & 1      & 1      & 1     & 1     & 2     & 1.2                   & 0.12                   \\ \cline{3-10} 
		\multicolumn{2}{|l|}{}                                    & $u_8$  & 1      & 1      & 1     & 1     & 1     & 1                     & 0.1                    \\ \cline{3-10} 
		\multicolumn{2}{|l|}{}                                    & $u_9$  & 1      & 1      & 1     & 1     & 1     & 1                     & 0.1                    \\ \hline
		\multirow{31}{*}{\rotatebox{90}{Secondary Indexes}} & \multirow{2}{*}{$u_1$} & $u_{11}$ & 5      & 5      & 5     & 4     & 4     & 4.6                   & 0.46                   \\ \cline{3-10} 
		&                     & $u_{12}$ & 5      & 5      & 5     & 6     & 6     & 5.4                   & 0.54                   \\ \cline{2-10} 
		& \multirow{5}{*}{$u_2$} & $u_{21}$ & 2      & 1      & 1     & 2     & 1     & 1.4                   & 0.14                   \\ \cline{3-10} 
		&                     & $u_{22}$ & 3      & 2      & 2     & 2     & 2     & 2.2                   & 0.22                   \\ \cline{3-10} 
		&                     & $u_{23}$ & 1      & 2      & 2     & 2     & 2     & 1.8                   & 0.18                   \\ \cline{3-10} 
		&                     & $u_{24}$ & 2      & 2      & 2     & 2     & 2     & 2                     & 0.2                    \\ \cline{3-10} 
		&                     & $u_{25}$ & 2      & 3      & 3     & 2     & 3     & 2.6                   & 0.26                   \\ \cline{2-10} 
		& \multirow{3}{*}{$u_3$} & $u_{31}$ & 3      & 3      & 3     & 4     & 3     & 3.2                   & 0.32                   \\ \cline{3-10} 
		&                     & $u_{32}$ & 3      & 4      & 3     & 4     & 3     & 3.4                   & 0.34                   \\ \cline{3-10} 
		&                     & $u_{33}$ & 4      & 3      & 4     & 2     & 4     & 3.4                   & 0.34                   \\ \cline{2-10} 
		& \multirow{4}{*}{$u_4$} & $u_{41}$ & 3      & 4      & 2     & 3     & 4     & 3.2                   & 0.32                   \\ \cline{3-10} 
		&                     & $u_{42}$ & 3      & 2      & 2     & 1     & 2     & 2                     & 0.2                    \\ \cline{3-10} 
		&                     & $u_{43}$ & 2      & 3      & 2     & 4     & 2     & 2.6                   & 0.26                   \\ \cline{3-10} 
		&                     & $u_{44}$ & 2      & 1      & 4     & 2     & 2     & 2.2                   & 0.22                   \\ \cline{2-10} 
		& \multirow{5}{*}{$u_5$} & $u_{51}$ & 1      & 2      & 2     & 1     & 2     & 1.6                   & 0.16                   \\ \cline{3-10} 
		&                     & $u_{52}$ & 2      & 2      & 2     & 2     & 2     & 2                     & 0.2                    \\ \cline{3-10} 
		&                     & $u_{53}$ & 2      & 2      & 1     & 2     & 2     & 1.8                   & 0.18                   \\ \cline{3-10} 
		&                     & $u_{54}$ & 3      & 2      & 3     & 2     & 2     & 2.4                   & 0.24                   \\ \cline{3-10} 
		&                     & $u_{55}$ & 2      & 2      & 2     & 3     & 2     & 2.2                   & 0.22                   \\ \cline{2-10} 
		& \multirow{4}{*}{$u_6$} & $u_{61}$ & 3      & 4      & 3     & 4     & 3     & 3.4                   & 0.34                   \\ \cline{3-10} 
		&                     & $u_{62}$ & 1      & 2      & 3     & 2     & 2     & 2                     & 0.2                    \\ \cline{3-10} 
		&                     & $u_{63}$ & 4      & 2      & 2     & 3     & 3     & 2.8                   & 0.28                   \\ \cline{3-10} 
		&                     & $u_{64}$ & 2      & 2      & 2     & 1     & 2     & 1.8                   & 0.18                   \\ \cline{2-10} 
		& \multirow{4}{*}{$u_7$} & $u_{71}$ & 3      & 4      & 3     & 3     & 4     & 3.4                   & 0.34                   \\ \cline{3-10} 
		&                     & $u_{72}$ & 3      & 2      & 2     & 2     & 1     & 2                     & 0.2                    \\ \cline{3-10} 
		&                     & $u_{73}$ & 2      & 1      & 3     & 3     & 2     & 2.2                   & 0.22                   \\ \cline{3-10} 
		&                     & $u_{74}$ & 2      & 3      & 2     & 2     & 3     & 2.4                   & 0.24                   \\ \cline{2-10} 
		& \multirow{2}{*}{$u_8$} & $u_{81}$ & 5      & 5      & 6     & 5     & 6     & 5.4                   & 0.54                   \\ \cline{3-10} 
		&                     & $u_{82}$ & 5      & 5      & 4     & 5     & 4     & 4.6                   & 0.46                   \\ \cline{2-10} 
		& \multirow{2}{*}{$u_9$} & $u_{91}$ & 5      & 5      & 5     & 4     & 5     & 4.8                   & 0.48                   \\ \cline{3-10} 
		&                     & $u_{92}$ & 5      & 5      & 5     & 6     & 5     & 5.2                   & 0.52                   \\ \hline
	\end{tabular}
\end{table}
\section{Evaluation process}
\label{evaluation process}
\subsection{Determination of weights}
The determination of each index's weight is one of the difficult tasks in multi-objective decision-making. Index weight refers to the index's value in the overall evaluation, the degree of relative importance, and the quantitative value of the proportion. In this paper, the weight represents the overall impact of the index in the evaluation of the security video surveillance system. The scientificity and rationality of each index's weight should be ensured. Therefore, we seek advice from 15 professional consultants to determine the weight of each index. Among the 15 consultants, we randomly choose the feedback from 5 of them as valid feedback. 

Each index's weight in the main evaluation system is determined according to its relative importance in the main system. Each index's weight in each subsystem is determined according to its relative importance in the same subsystem. In the main system, the weight of each index is from 0 to 10, and the sum of each index is equal to 10. In each subsystem, the weight of each index is from 0 to 10, and the sum of each index in the same subsystem is equal to 10. Each index's final weight is the average value of 5 valid feedback, and then it is normalized. All index weights are calculated and shown in table \ref{Weight table}. 
\begin{table}[t]
	\caption{The membership table}
	\label{the membership table}
	\centering
	\begin{tabular}{|l|l|l|l|l|l|}
		\hline
		\multirow{2}{*}{Indexes} & \multicolumn{5}{l|}{Membership}          \\ \cline{2-6} 
		& Excellent & Good & Average & Fair & Poor \\ \hline
		$u_{11}$                     & 0.7       & 0.2  & 0.1     & 0    & 0    \\ \hline
		$u_{12}$                    & 0.8       & 0.1  & 0.1     & 0    & 0    \\ \hline
		$u_{21}$                    & 0.9       & 0.1  & 0       & 0    & 0    \\ \hline
		$u_{22}$                     & 0.7       & 0.3  & 0       & 0    & 0    \\ \hline
		$u_{23}$                     & 0.8       & 0.2  & 0       & 0    & 0    \\ \hline
		$u_{24}$                     & 0.8       & 0.2  & 0       & 0    & 0    \\ \hline
		$u_{25}$                     & 0.7       & 0.2  & 0.1     & 0    & 0    \\ \hline
		$u_{31}$                     & 0.9       & 0.1  & 0       & 0    & 0    \\ \hline
		$u_{32}$                     & 0.7       & 0.2  & 0.1     & 0    & 0    \\ \hline
		$u_{33}$                     & 0.9       & 0.1  & 0       & 0    & 0    \\ \hline
		$u_{41}$                     & 0.8       & 0.1  & 0.1     & 0    & 0    \\ \hline
		$u_{42}$                     & 0.7       & 0.2  & 0.1     & 0    & 0    \\ \hline
		$u_{43}$                     & 0.8       & 0.1  & 0.1     & 0    & 0    \\ \hline
		$u_{44}$                     & 0.9       & 0.1  & 0       & 0    & 0    \\ \hline
		$u_{51}$                     & 0.8       & 0.1  & 0.1     & 0    & 0    \\ \hline
		$u_{52}$                     & 0.8       & 0.2  & 0       & 0    & 0    \\ \hline
		$u_{53}$                     & 0.9       & 0.1  & 0       & 0    & 0    \\ \hline
		$u_{54}$                     & 0.8       & 0.1  & 0.1     & 0    & 0    \\ \hline
		$u_{55}$                     & 0.7       & 0.2  & 0.1     & 0    & 0    \\ \hline
		$u_{61}$                     & 0.8       & 0.2  & 0       & 0    & 0    \\ \hline
		$u_{62}$                     & 0.7       & 0.3  & 0       & 0    & 0    \\ \hline
		$u_{63}$                     & 0.7       & 0.2  & 0.1     & 0    & 0    \\ \hline
		$u_{64}$                     & 0.9       & 0.1  & 0       & 0    & 0    \\ \hline
		$u_{71}$                     & 0.8       & 0.2  & 0       & 0    & 0    \\ \hline
		$u_{72}$                     & 0.8       & 0.1  & 0.1     & 0    & 0    \\ \hline
		$u_{73}$                     & 0.9       & 0.1  & 0       & 0    & 0    \\ \hline
		$u_{74}$                     & 0.7       & 0.3  & 0       & 0    & 0    \\ \hline
		$u_{81}$                     & 0.8       & 0.1  & 0.1        & 0    & 0    \\ \hline
		$u_{82}$                   & 0.9       & 0.1  & 0       & 0    & 0    \\ \hline
		$u_{91}$                    & 0.7       & 0.2  & 0.1     & 0    & 0    \\ \hline
		$u_{92}$                     & 0.8       & 0.1  & 0.1     & 0    & 0    \\ \hline
	\end{tabular}
\end{table}
\subsection{Determination of membership}
The determination of membership is essential in multi-level fuzzy evaluation. We seek advice from 15 professional consultants to determine the membership of each secondary index. 
For each $u_i\in U'$, each of the 15 consultants provides a comment $v_j\in V$ according to the consultant's view of the implementation of the surveillance system regarding the index $u_i$.
For index $u_{91}$ and $u_{92}$, the consultants provide comments based on the survey results.
Among the 15 consultants, we randomly choose the feedback from 10 of them as valid feedback. The following formula is used to calculate the membership of each secondary index $u_i$:
\begin{equation}
M_{ij}=m_{ij}/n
\end{equation}
$M_{ij}$ is the membership of index $u_i$ that commented as $v_j$; $m_{ij}$ is the number of valid feedback of index $u_i$ that commented as $v_j$; $n$ is the total number of valid feedback (10 in this case). The memberships of all secondary indexes are calculated and shown in table \ref{the membership table}.

\subsection{Determination of evaluation matrices}
From table \ref{the membership table}, the evaluation matrices $r_1$\textasciitilde$r_9$ corresponding to $u_1$\textasciitilde$u_9$ can be derived as follows:
\begin{flalign}
\nonumber
&\left\{
\begin{aligned}
&{r}_{1}=\left(\begin{array}{ccccc}
0.7 & 0.2 & 0.1 & 0 & 0 \\
0.8 & 0.1 & 0.1 & 0 & 0 
\end{array}\right)\\
&{r}_{2}=\left(\begin{array}{ccccc}
0.9       & 0.1  & 0       & 0    & 0    \\ 
0.7       & 0.3  & 0       & 0    & 0    \\ 
0.8       & 0.2  & 0       & 0    & 0    \\ 
0.8       & 0.2  & 0       & 0    & 0    \\ 
0.7       & 0.2  & 0.1     & 0    & 0
\end{array}\right)\\
&{r}_{3}=\left(\begin{array}{ccccc}
0.9       & 0.1  & 0       & 0    & 0    \\ 
0.7       & 0.2  & 0.1     & 0    & 0    \\ 
0.9       & 0.1  & 0       & 0    & 0    
\end{array}\right)\\
&{r}_{4}=\left(\begin{array}{ccccc}
0.8       & 0.1  & 0.1     & 0    & 0    \\ 
0.7       & 0.2  & 0.1     & 0    & 0    \\ 
0.8       & 0.1  & 0.1     & 0    & 0    \\ 
0.9       & 0.1  & 0       & 0    & 0    \\  
\end{array}\right)\\
&{r}_{5}=\left(\begin{array}{ccccc}
0.8       & 0.1  & 0.1     & 0    & 0    \\ 
0.8       & 0.2  & 0       & 0    & 0    \\ 
0.9       & 0.1  & 0       & 0    & 0    \\ 
0.8       & 0.1  & 0.1     & 0    & 0    \\ 
0.7       & 0.2  & 0.1     & 0    & 0    \\ 
\end{array}\right)\\
&{r}_{6}=\left(\begin{array}{ccccc}
0.8       & 0.2  & 0       & 0    & 0    \\  
0.7       & 0.3  & 0       & 0    & 0    \\  
0.7       & 0.2  & 0.1     & 0    & 0    \\  
0.9       & 0.1  & 0       & 0    & 0    \\ 
\end{array}\right)\\
&{r}_{7}=\left(\begin{array}{ccccc}
0.8       & 0.2  & 0       & 0    & 0    \\ 
0.8       & 0.1  & 0.1     & 0    & 0    \\ 
0.9       & 0.1  & 0       & 0    & 0    \\ 
0.7       & 0.3  & 0       & 0    & 0    \\
\end{array}\right)\\
&{r}_{8}=\left(\begin{array}{ccccc}
0.8       & 0.1  & 0.1        & 0    & 0    \\ 
0.9       & 0.1  & 0       & 0    & 0    \\ 
\end{array}\right)\\
&{r}_{9}=\left(\begin{array}{ccccc}
0.7       & 0.2  & 0.1     & 0    & 0    \\ 
0.8       & 0.1  & 0.1     & 0    & 0    \\ 
\end{array}\right)
\end{aligned}
\right.&
\end{flalign}

\subsection{The evaluation result}
From table \ref{Weight table}, the weight vectors $a_1$\textasciitilde$a_9$ corresponding to $u_1$\textasciitilde$u_9$ can be derived as follows:
\begin{flalign}
\nonumber
&\left\{
\begin{aligned}
&a_1 = (0.46\ 0.54)\\
&a_2 = (0.14\ 0.22\ 0.18\ 0.2\ 0.26)\\
&a_3 = (0.32\ 0.34\ 0.34)\\
&a_4 = (0.32\ 0.2\ 0.26\ 0.22)\\
&a_5 = (0.16\ 0.2\ 0.18\ 0.24\ 0.22)\\
&a_6 = (0.34\ 0.2\ 0.28\ 0.18)\\
&a_7 = (0.34\ 0.2\ 0.22\ 0.24)\\
&a_8 = (0.54\ 0.46)\\
&a_9 = (0.48\ 0.52)
\end{aligned}
\right.&
\end{flalign}
\indent According to the formula of fuzzy evaluation, the vectors of evaluation result $b_1$\textasciitilde$b_9$ corresponding to index $u_1$\textasciitilde$u_9$ can be calculated as follows:
\begin{flalign}\nonumber
&\left\{
\begin{aligned}
&b_1 = a_1r_1 = (0.754\ 0.146\ 0.1\ 0\ 0)\\
&b_2 = a_2r_2 = (0.766\ 0.208\ 0.026\ 0\ 0)\\
&b_3 = a_3r_3 = (0.832\ 0.134\ 0.034\ 0\ 0)\\
&b_4 = a_4r_4 = (0.802\ 0.12\ 0.078\ 0\ 0)\\
&b_5 = a_5r_5 = (0.796\ 0.142\  0.062\ 0\ 0)\\
&b_6 = a_6r_6 = (0.77\ 0.202\ 0.028\ 0\ 0)\\
&b_7 = a_7r_7 = (0.798\  0.182\   0.02\ 0\ 0)\\
&b_8 = a_8r_8 = (0.846\   0.1\   0.054\   0\ 0)\\
&b_9 = a_9r_9 = (0.752\   0.148\   0.1\   0\ 0)\\
\end{aligned}
\right.&
\end{flalign}
\indent From table \ref{Weight table}, the weight vector $A$ corresponding to $U$ can be derived as follows:
\begin{flalign}\nonumber
&A=(0.1\ 0.12\ 0.1\ 0.12\ 0.12\ 0.12\ 0.12\ 0.1\ 0.1)&
\end{flalign}

\indent According to the formula of fuzzy evaluation, the final evaluation vector $B$ is:
\begin{flalign}\nonumber
&B = A\left(\begin{array}{ccccc}
b_1   \\ 
b_2   \\ 
...\\
b_9   \\ 
\end{array}\right)=(0.79024\ 0.15528\ 0.05448\ 0\ 0)&
\end{flalign}

As $V=\{5, 4, 3, 2, 1\}$, according to the principle of the highest degree of membership (0.79024), the evaluation result is "Excellent (Level 5)". It shows that the construction and implementation of the video surveillance system have reached an excellent level. Therefore, the evaluation result demonstrates that the proposed approach is practically applicable to real-world security video surveillance systems.
\section{Conclusion and future work}
\label{Conclusion}
This paper aims to provide a feasible approach to evaluate the construction and implementation of a typical real-world video surveillance system. The fuzzy theory and fuzzy multi-level evaluation method are applied in the proposed approach. 
However, currently, there is a limited number of sound evaluation index systems for evaluating video surveillance systems. The methods for the determination of index weights also need to be improved.
We will gradually improve the index system and the methods for the determination of index weights in the future. We will try to use artificial intelligence, knowledge engineering, expert system, and other theoretical approaches to establish a comprehensive evaluation system based on knowledge. The system should be of universality and scalability to provide practical support for new post-project evaluation problems.

\ifCLASSOPTIONcaptionsoff
\newpage
\fi
\bibliographystyle{IEEEtran}
\bibliography{mybibfile}

\begin{thebibliography}{1}
\providecommand{\url}[1]{#1}
\csname url@samestyle\endcsname
\providecommand{\newblock}{\relax}
\providecommand{\bibinfo}[2]{#2}
\providecommand{\BIBentrySTDinterwordspacing}{\spaceskip=0pt\relax}
\providecommand{\BIBentryALTinterwordstretchfactor}{4}
\providecommand{\BIBentryALTinterwordspacing}{\spaceskip=\fontdimen2\font plus
\BIBentryALTinterwordstretchfactor\fontdimen3\font minus
  \fontdimen4\font\relax}
\providecommand{\BIBforeignlanguage}[2]{{%
\expandafter\ifx\csname l@#1\endcsname\relax
\typeout{** WARNING: IEEEtran.bst: No hyphenation pattern has been}%
\typeout{** loaded for the language `#1'. Using the pattern for}%
\typeout{** the default language instead.}%
\else
\language=\csname l@#1\endcsname
\fi
#2}}
\providecommand{\BIBdecl}{\relax}
\BIBdecl

\bibitem{Collins2000}
R.~Collins, A.~Lipton, T.~Kanade, H.~Fujiyoshi, D.~Duggins, Y.~Tsin,
  D.~Tolliver, N.~Enomoto, O.~Hasegawa, and P.~Burt, ``A system for video
  surveillance and monitoring,'' \emph{Robot. Inst.}, vol.~5, 06 2000.

\bibitem{Vishwakarma2013}
S.~Vishwakarma and A.~Agrawal, ``A survey on activity recognition and behavior
  understanding in video surveillance,'' \emph{The Visual Computer}, vol.~29,
  no.~10, pp. 983--1009, 2013.

\bibitem{Ojha2015}
S.~{Ojha} and S.~{Sakhare}, ``Image processing techniques for object tracking
  in video surveillance-a survey,'' in \emph{2015 International Conference on
  Pervasive Computing (ICPC)}, 2015, pp. 1--6.

\bibitem{TDraty2010}
T.~D. {Raty}, ``Survey on contemporary remote surveillance systems for public
  safety,'' \emph{IEEE Transactions on Systems, Man, and Cybernetics, Part C
  (Applications and Reviews)}, vol.~40, no.~5, pp. 493--515, 2010.

\bibitem{Anbari2008Post}
F.~T. Anbari, E.~G. Carayannis, and R.~J. Voetsch, ``Post-project reviews as a
  key project management competence,'' \emph{Technovation}, vol.~28, no.~10,
  pp. 0--643, 2008.

\bibitem{wangyongsheng}
Y.~Wang and G.~Xing, ``Research on evaluation index system of public security
  video monitoring network application,'' \emph{China Security and Protection},
  vol.~4, pp. 96--101, 2018.

\end{thebibliography}
\end{document}